\documentclass[12pt,a4paper,fleqn]{article}

\usepackage[DIV12]{typearea}
\usepackage{amsmath}
\usepackage{amssymb}
\usepackage{amsfonts}
\usepackage{amscd}
\usepackage{graphicx}
\usepackage[footnotesize]{caption2}
\usepackage{mcite}
\usepackage{mathrsfs}

\newcommand{\GeV}{\ensuremath{\:\mathrm{GeV}}}
\newcommand{\TeV}{\ensuremath{\:\mathrm{TeV}}}

\newcommand{\Eqref}[1]{Eq.~\eqref{#1}}
\newcommand{\Figref}[1]{Fig.~\ref{#1}}

\def\D{\mathrm{d}}
\def\I{i}

\allowdisplaybreaks[1]

\begin{document}

\begin{titlepage}
\renewcommand{\thefootnote}{\alph{footnote}}

\begin{flushright}
DESY 05-089
\end{flushright}

\vspace*{1.0cm}

\renewcommand{\thefootnote}{\fnsymbol{footnote}}

\begin{center}
{\Large\bf The Gravitino in Gaugino Mediation}

\vspace*{1cm}
\renewcommand{\thefootnote}{\alph{footnote}}

\textbf{
Wilfried Buchm\"uller\footnote[1]{Email: \texttt{wilfried.buchmueller@desy.de}},
Koichi Hamaguchi\footnote[2]{Email: \texttt{koichi.hamaguchi@desy.de}}
and
J\"orn Kersten\footnote[3]{Email: \texttt{joern.kersten@desy.de}}
}
\\[5mm]

Deutsches Elektronen-Synchrotron DESY, 22603 Hamburg, Germany
\end{center}

\vspace*{1cm}

\begin{abstract}
\noindent
We show that for gaugino mediated supersymmetry breaking the gravitino
mass is bounded from below.  For a size of the compact dimensions of
order the unification scale and a cutoff given by the
higher-dimensional Planck mass, we find $m_{3/2} \gtrsim 10\GeV$.  In a
large domain of parameter space, the gravitino is the lightest
superparticle with a scalar $\tilde\tau$-lepton as the next-to-lightest
superparticle.

\end{abstract}

\end{titlepage}

\newpage

\section{Introduction}
In supersymmetric (SUSY) models with R-parity conserved, the lightest
superparticle (LSP) is stable and plays a key role in cosmology as
well as in collider physics. The most widely studied LSP candidate is
the neutralino, which is a linear combination of the gauginos and
higgsinos. There is however another candidate, the gravitino, which is always
present once SUSY is extended to a local symmetry leading to supergravity.

As a neutral and stable particle, the gravitino LSP is a natural
cold dark matter candidate~\cite{Pagels:1981ke}.\footnote{A keV
gravitino as dominant component of dark matter as discussed in
\cite{Pagels:1981ke} is now disfavored by the matter power spectrum
(cf.\ \cite{Viel:2005qj}).}  In the early universe gravitinos
are produced by thermal scatterings after
inflation~\cite{Nanopoulos:1983up,Khlopov:1984pf,*Ellis:1984eq,Moroi:1993mb}. Generically,
their relic abundance exceeds the observed cold dark
matter density unless the reheating temperature $T_R$ is sufficiently
small.  However, several mechanisms have been proposed 
which avoid this constraint and yield the correct dark matter
density even for large $T_R$
\cite{Bolz:1998ek,*Fujii:2002fv,*Fujii:2003iw,Buchmuller:2003is}. 
This also removes an obstacle for thermal leptogenesis \cite{Fukugita:1986hr}, 
which requires $T_R \gtrsim 2 \cdot 10^9\GeV$~\cite{Buchmuller:2002rq}.
Such a high reheating temperature is disfavored for a non-LSP gravitino,
since its decays would alter the light element abundances produced by
big-bang nucleosynthesis~\cite{Falomkin:1984eu,Khlopov:1984pf,*Ellis:1984eq,Kawasaki:2004qu}.
Even for a gravitino LSP, the late time decay of the next-to-lightest
superparticle (NLSP) may cause
cosmological problems.  Recent analyses show that a stau
NLSP is allowed for a gravitino mass $\lesssim 10\,\text{--}\,100\GeV$,
while for a neutralino NLSP the bound is severer
\cite{Fujii:2003nr,*Ellis:2003dn,*Feng:2004mt,*Roszkowski:2004jd}.
Gravitino dark matter can also be realized via non-thermal
production from decays of the NLSP~\cite{Feng:2003xh,*Feng:2003uy}.

If the long-lived NLSP is a charged particle, such as the stau, it can
be collected and studied in detail at the LHC and the ILC, which allows
to explore various aspects of new
physics~\cite{Buchmuller:2004rq,Hamaguchi:2004df,*Feng:2004yi,*Hamaguchi:2004ne,*Brandenburg:2005he}. Particularly
interesting are gravitino masses in the range $10\,$--$\,100\GeV$, where one
may be able to measure the Planck scale as well as
the gravitino spin~\cite{Buchmuller:2004rq}.

The theoretical predictions for the gravitino mass and the nature of the
LSP depend on the mechanism of SUSY breaking.  In models with gauge
mediation, the gravitino is usually much lighter than
$1\GeV$~\cite{Dine:1994vc,*Dine:1995ag}, so that it clearly becomes the
LSP.  For gravity mediation, its mass is of the same order as the masses
of scalar quarks and leptons, i.e.\ $100\GeV\,$--$\,1\TeV$
\cite{Nilles:1983ge}.  Whether it is the LSP or not depends on the
details of the model.  On the other hand, anomaly mediation predicts a
very heavy gravitino \cite{Randall:1998uk,*Giudice:1998xp}, which cannot
be the LSP.

In this Letter we discuss the gravitino mass for gaugino mediated SUSY
breaking~\cite{Kaplan:1999ac,*Chacko:1999mi}, which is one of the
simplest mechanisms solving the SUSY flavor problem (cf.\ also
\cite{Inoue:1991rk}).  The role of the
gravitino in this context has not been studied in detail in the
literature.  We use naive dimensional analysis (NDA) to derive a lower
bound on the gravitino to gaugino mass ratio.  From this we conclude
that the gravitino mass is typically larger than about $10\GeV$.
Therefore, it can be the LSP.  In this case the NLSP is naturally the
stau.  Together with the relatively large gravitino mass, this has
exciting consequences for cosmology and collider physics.

\section{Gaugino Mediation}
We consider a theory with $D$ dimensions and 4-dimensional branes
located at positions $y_i$ in the compact dimensions.
Coordinates $x$ denote the usual 4 dimensions, while $y$ refer to the
compact dimensions.  
In models with gaugino mediated SUSY breaking \cite{Kaplan:1999ac,*Chacko:1999mi}, the gauge superfields live in the
bulk, while the scalar responsible for SUSY breaking
(contained in the chiral superfield $S$) lives on the 4-dimensional
brane $i=1$.  The part of the Lagrangian relevant for gaugino masses is
\begin{align} \label{eq:LDOriginal}
	\mathscr{L}_D &=
	\frac{1}{4g_D^2} \int\D^2\theta \, W^a W^a + \text{h.c.} + {}
\nonumber\\
& \quad\; +
	\delta^{(D-4)}(y-y_1) \int\D^4\theta \, S^\dagger S + {}
\nonumber\\
& \quad\; +
	\delta^{(D-4)}(y-y_1)\,\frac{h}{4\Lambda} \int\D^2\theta\,S\,W^a W^a
	+ \text{h.c.} \;,
\end{align}
where $W^a$ is the field strength superfield, $h$ is a dimensionless
coupling and $\Lambda$ is the cutoff of
the theory.  All fields are 4D $N=1$ superfields.  Bulk fields depend on
the coordinates $y$.  For details of the formalism, see
\cite{Arkani-Hamed:2001tb,*Hebecker:2001ke}.  Additional fields required
by the higher-dimensional SUSY are present but not explicitly included
in the Lagrangian, since they are not relevant for our discussion.

A vacuum expectation value (vev) $F_S$ for the $F$-term of $S$ breaks SUSY
and leads to the gaugino mass
\begin{equation} \label{eq:GauginoMass}
	m_{1/2} = \frac{g_4^2 h \, F_S}{2\Lambda}
\end{equation}
at the compactification scale.
The gravitino mass is given by \cite{Nilles:1983ge}
\begin{equation} \label{eq:GravitinoMass}
	m_{3/2} = \frac{1}{\sqrt{3}} \frac{F_S}{M_4} \;,
\end{equation}
where $M_4 \simeq 2.4 \cdot 10^{18} \GeV$ is the 4-dimensional
(reduced) Planck mass.  This relation is valid if the vev of $F_S$ is
the only source of SUSY breaking.  If there are further sources, the
gravitino becomes heavier.

\section{Constraints from Naive Dimensional Analysis}

We want the effective Lagrangian (\ref{eq:LDOriginal}) to be valid up to
a cutoff scale $\Lambda$. This requires that the couplings at the
compactification scale do not exceed
upper bounds which can be estimated by means of `naive dimensional
analysis' \cite{Chacko:1999hg}.  In general, one rewrites the
$D$-dimensional Lagrangian with bulk 
fields $\Phi(x,y)$ and brane fields $\phi_i(x)$ on the $i$th brane, 
\begin{equation} \label{eq:LDCanonical}
	\mathscr{L}_D = 
	\mathscr{L}_\mathrm{bulk}(\Phi(x,y)) +
	\sum_i \delta^{D-4}(y-y_i) \, \mathscr{L}_i(\Phi(x,y),\phi_i(x)) \;,
\end{equation}
in terms of dimensionless fields $\hat\Phi(x,y)$ and $\hat\phi_i(x)$,
and the cutoff $\Lambda$, so that
\begin{equation} \label{eq:LDDimless}
	\mathscr{L}_D = 
	\frac{\Lambda^D}{\ell_D/C} \,
	 \mathscr{\hat L}_\mathrm{bulk}(\hat\Phi(x,y)) +
	\sum_i \delta^{D-4}(y-y_i) \, \frac{\Lambda^4}{\ell_4/C} \,
	 \mathscr{\hat L}_i(\hat\Phi(x,y),\hat\phi_i(x)) \;.
\end{equation}
Here the Lagrangians $\mathscr{\hat L}$ have kinetic terms of the form
$\mathscr{\hat L} = (\frac{\partial}{\Lambda}\hat\Phi)^2 + \dots$ for
scalars, and analogously for other fields.  If the kinetic terms of the
original Lagrangian \eqref{eq:LDCanonical} are canonical with respect to
$\Phi$ and $\phi_i$, the rescaling of bosonic bulk and brane fields reads  
\begin{equation}
\Phi(x,y) = \left( \frac{\Lambda^{D-2}}{\ell_D/C} \right)^{1/2}\hat\Phi(x,y)
\quad , \quad
\phi_i(x) = \left( \frac{\Lambda^2}{\ell_4/C} \right)^{1/2} \hat\phi_i(x) \;.
\label{eq:phiAndphiHat}
\end{equation}
For non-canonical kinetic terms in \Eqref{eq:LDCanonical}, the field
rescaling has to be adjusted so that \Eqref{eq:LDDimless} is obtained.
The geometrical loop factor
\begin{equation}
	\ell_D = 2^D \pi^{D/2} \, \Gamma(D/2)
\end{equation}
grows rapidly with the number of dimensions: $\ell_4 = 16\pi^2$,
$\ell_5 = 24\pi^3$, $\ell_6 = 128\pi^3$ etc. The factor $C$
accounts for the multiplicity of fields in loop diagrams for a
non-Abelian gauge group $G$.
We choose $C=C_2(G)$, i.e.\ $C=5$ for SU(5) and $C=8$ for SO(10).  

The combination $C/\ell_D$ gives the typical geometrical suppression of
loop diagrams.  This suppression is canceled by the factors
$\ell_D/C$ and $\ell_4/C$ in front of the Lagrangians $\mathscr{\hat L}$
in \Eqref{eq:LDDimless}.  Consequently, all loops will be of the
same order of magnitude, provided that all couplings are
$\mathscr{O}(1)$.
Thus, according to the NDA recipe the effective $D$-dimensional theory remains weakly
coupled up to the cutoff $\Lambda$, if the dimensionless couplings in  
\Eqref{eq:LDDimless} are smaller than one. 

As an example, consider the $D$-dimensional gauge coupling
\begin{equation}\label{gaugeD4}
	\frac{V_{D-4}}{g_D^2} = \frac{1}{g_4^2} \;,
\end{equation}
where $V_{D-4}$ is the volume of the compact dimensions.
From the dimensionless covariant derivative, 
\begin{equation}
	\hat D_\mu =  
	\frac{\partial_\mu}{\Lambda} - \frac{\I g_D A_\mu}{\Lambda} =
	\frac{\partial_\mu}{\Lambda} -
	 \I g_D \left(\frac{\Lambda^{D-4}}{\ell_D/C}\right)^{1/2} \hat A_\mu
	\;,
\end{equation}
one reads off
\begin{equation}
	g_D \left(\frac{\Lambda^{D-4}}{\ell_D/C}\right)^{1/2} < 1 \;.
\end{equation}
For a given cutoff this constrains the gauge coupling. Conversely, knowing
the gauge coupling at the compactification scale, $g_4^2$, 
one obtains an upper bound on the cutoff,
\begin{equation}
	\Lambda < \Lambda_\mathrm{gauge} =
	\left( \frac{\ell_D/C}{g_4^2} \right)^\frac{1}{D-4} M_c \;,
\end{equation}
where we have defined
\begin{equation} \label{uniV}
	M_c = \left( \frac{1}{V_{D-4}} \right)^\frac{1}{D-4} \;.
\end{equation}
For $M_c$ close to the unification scale, one has
$g_4^2\simeq\frac{1}{2}$.

\begin{figure}
\centering
\includegraphics{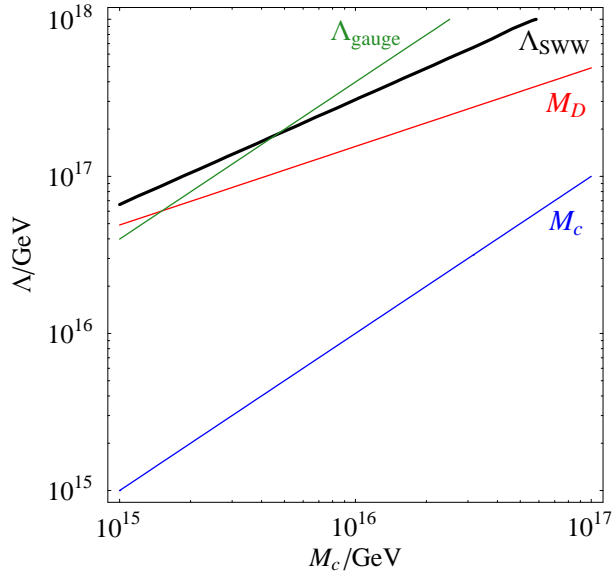}
\caption{Relevant scales ($D=6$): $\Lambda_{SWW}$ for $m_{1/2}=m_{3/2}$ and
 $C=5$, ignoring the running of the gaugino masses,
 $\Lambda_\mathrm{gauge}$ for $g_4^2=\frac{1}{2}$, $D$-dimensional Planck scale,
 and the lower limit $\Lambda>M_c$.
}
\label{fig:Scales}
\end{figure}

The cutoff $\Lambda_\mathrm{gauge}$ can be compared with the 
$D$-dimensional Planck scale $M_D$ where quantum gravity effects are expected
to become important,
\begin{equation}
	M_D = \left( M_c^{D-4} M_4^2 \right)^\frac{1}{D-2} \;,
\end{equation}
as shown in \Figref{fig:Scales} for $D=6$.  We require $\Lambda<M_D$,
which turns out to be more restrictive than 
$\Lambda < \Lambda_\mathrm{gauge}$, unless $M_c$ is very small.

\section{Application to Gaugino Mediation}

We can now apply the NDA prescription to gaugino mediation, which will lead to
an upper bound on the gaugino masses.
The field strength superfield has to be rescaled as
(cf.\ \Eqref{eq:LDOriginal} and \eqref{eq:LDDimless})
\begin{equation} \label{eq:WAndWHat}
	W^a = \left( \frac{\Lambda^{D-1}}{\ell_D/C} \right)^{1/2} g_D \,
	\hat W^a \;.
\end{equation}
Since $\D\theta$ has mass dimension $1/2$, it also has to be divided
by the corresponding power of the cutoff to obtain a dimensionless expression.
We then arrive at the Lagrangian
\begin{align} \label{eq:LGauginoDimless}
	\mathscr{L}_D &=
	\frac{\Lambda^D}{\ell_D/C} \, \frac{1}{4} 
	\int\frac{\D^2\theta}{\Lambda} \,\hat W^a \hat W^a + \text{h.c.} +{}
\nonumber\\
& \quad\; +
	\delta^{(D-4)}(y-y_1) \, \frac{\Lambda^4}{\ell_4/C}
	\int\frac{\D^4\theta}{\Lambda^2} \, \hat S^\dagger \hat S + {}
\nonumber\\
& \quad\; +
	\delta^{(D-4)}(y-y_1) \, \frac{\Lambda^4}{\ell_4/C} \,
	\frac{g_D^2 h \sqrt{\ell_4 C} \Lambda^{D-4}}{\ell_D} \, \frac{1}{4}
	\int\frac{\D^2\theta}{\Lambda} \, \hat S \, \hat W^a \hat W^a
	+ \text{h.c.}
\end{align}
The requirement that all couplings%
\footnote{Note that this applies to the couplings of canonically
 normalized fields.  Hence, the factor $\frac{1}{4}$ in the last line
 of \Eqref{eq:LGauginoDimless} is not part of the coupling.
}
be smaller than one implies
\begin{equation}
	\frac{g_D^2 h \sqrt{\ell_4 C} \Lambda^{D-4}}{\ell_D} < 1 \;.
\end{equation}
Using the relations (\ref{gaugeD4}) and (\ref{uniV}), and $\ell_4=16\pi^2$,
one then obtains an upper bound on the coupling $h$,
\begin{equation}
	h < \frac{\ell_D}{4\pi\sqrt{C} g_4^2}
	\left( \frac{M_c}{\Lambda} \right)^{D-4} ,
\end{equation}
which translates into an upper bound on the gaugino mass (cf.\
\Eqref{eq:GauginoMass}):
\begin{equation} \label{eq:UpperBoundGauginoMass}
	m_{1/2} <  
	\frac{\ell_D F_S}{8\pi \sqrt{C} \Lambda}
	\left( \frac{M_c}{\Lambda} \right)^{D-4} .
\end{equation}
Note that there is no lower bound on the gaugino mass. The upper bound
becomes weaker if the cutoff $\Lambda$ is lowered.  Together with
\Eqref{eq:GravitinoMass}, \Eqref{eq:UpperBoundGauginoMass} yields a
lower bound on the mass ratio
\begin{equation} \label{eq:LowerBoundMassRatio}
	\frac{m_{3/2}}{m_{1/2}} >
	\frac{8\pi \sqrt{C}}{\sqrt{3} \ell_D}
	\left( \frac{\Lambda}{M_c} \right)^{D-4} \frac{\Lambda}{M_4} \;.
\end{equation}
For a fixed gravitino to gaugino mass ratio, \Eqref{eq:LowerBoundMassRatio}
yields again an upper bound on the cutoff $\Lambda$,
\begin{equation} \label{eq:LambdaSWW}
	\Lambda < \Lambda_{SWW} =
	\left( \frac{\sqrt{3} \ell_D}{8\pi \sqrt{C}} \, M_4 \, M_c^{D-4} \
	\frac{m_{3/2}}{m_{1/2}}
	\right)^\frac{1}{D-3} \;,
\end{equation}
which is compared in \Figref{fig:Scales} with $\Lambda_\mathrm{gauge}$
and $M_D$ as a function of $M_c$.   

\begin{figure}
\centering
\includegraphics{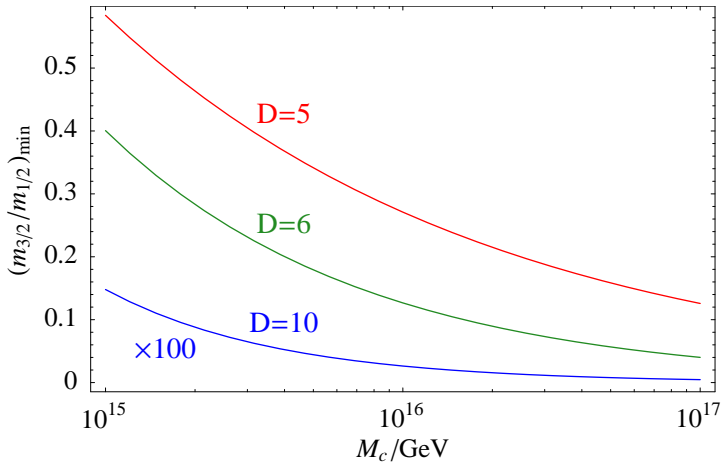}
\caption{Lower bound on the ratio of the gravitino and the gaugino mass
 (valid at the compactification scale) for a cutoff $\Lambda=M_D$,
 $C=5$, and $D=5,6,10$.  The $D=10$ result is multiplied by a factor of
 100.
}
\label{fig:RatioM12M32}
\end{figure}

Let us now discuss the lower bound \eqref{eq:LowerBoundMassRatio}
on the gravitino to gaugino mass ratio for a given cutoff $\Lambda$.
For the minimal value $\Lambda=M_c$, one obtains the absolute lower
bound $m_{3/2}/m_{1/2} > 8\pi \sqrt{C}/(\sqrt{3} \ell_D) \, M_c/M_4$.
However, this corresponds to the extreme case where the effective theory
described by the Lagrangian \eqref{eq:LDOriginal} becomes
non-perturbative immediately above $M_c$.  In the following, we choose
the cutoff to equal the $D$-dimensional Planck scale for concreteness.
The minimal mass ratio is then a function of the number of dimensions
and the compactification scale,
\begin{equation}
	\left( \frac{m_{3/2}}{m_{1/2}} \right)_\mathrm{min} =
	\frac{8\pi \sqrt{C}}{\sqrt{3} \ell_D}
	\left( \frac{M_4}{M_c} \right)^\frac{D-4}{D-2} \;.
\end{equation}
It is shown
in \Figref{fig:RatioM12M32} for $D=5,6,10$ and compactification scales
between $10^{15} \GeV$ and $10^{17} \GeV$.  As we are assuming compact
dimensions of equal size, the $D=10$ example appears less favored
\cite{Hebecker:2004ce} and should only be considered as a limiting case
for illustration purposes.
Note that here $m_{1/2}$ is the value of the gaugino mass at the
compactification scale.  The running to low energies typically decreases
the mass of the lightest gaugino by a factor of about $0.4$.  We have
not included this correction here, since it is model-dependent.
From the figure we see that the gravitino can be the LSP, if the gaugino
mass is sufficiently close to its upper bound from NDA.  However, it
cannot be much lighter than the neutralinos for $D=5$ and $D=6$.
This also means that a gravitino LSP becomes unlikely if the
theory is only weakly coupled at the cutoff.

\begin{figure}
\centering
\includegraphics{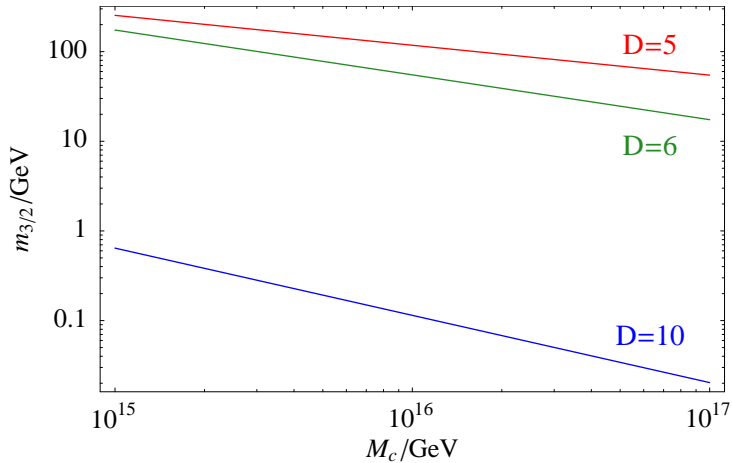}
\caption{Lower bound on the gravitino mass for
 $M_3(1\TeV) = 1\TeV$, $\Lambda=M_D$, $C=5$, and $D=5,6,10$.
}
\label{fig:GravitinoMass}
\end{figure}

Fixing $m_{1/2}$, we find a lower bound for $m_{3/2}$, which is shown in
\Figref{fig:GravitinoMass} for a gluino mass (at low energy)
of $1\TeV$.  For $D=6$ and $M_c=10^{17}\GeV$, we obtain
$m_{3/2}>17\GeV$.  This can be considered a typical lower bound in
gaugino mediation.  Hence, experimental implications for a much lighter
gravitino would disfavor this mechanism of SUSY breaking.
To realize smaller gravitino masses, one can lower $m_{1/2}$
or the cutoff scale, or increase the number of extra dimensions.  
Further possibilities include the group theory factor $C$, which could
be smaller if the GUT symmetry was broken on the brane where $S$ is
located.

\begin{figure}
\centering
\includegraphics{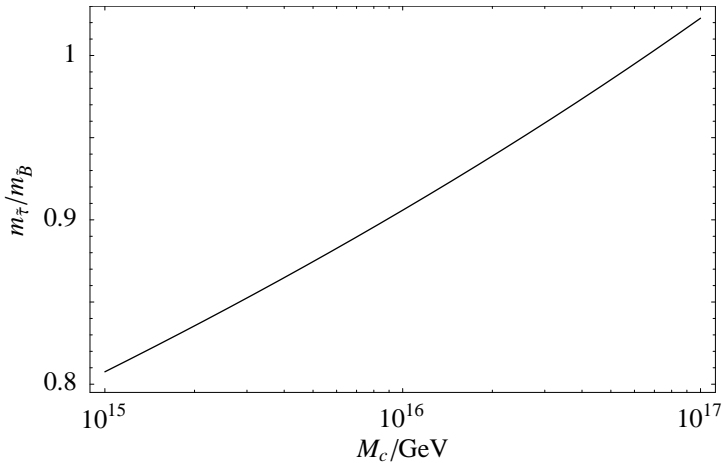}
\caption{Ratio of stau and bino mass for $\tan\beta=10$ and
 $m_{1/2}(M_c) = 400\GeV$, calculated in a very rough approximation (see
 text).
}
\label{fig:RatioMStauMBino}
\end{figure}

Another important aspect of the scenario is the mass of the lighter
stau.  In gaugino mediation, it is very small at the compactification
scale, so that its dominant contribution is due to the running to low
energies.  It is determined by the renormalization group equations of
the right-handed stau mass squared \cite{Inoue:1982pi,*Inoue:1983pp},
\begin{equation} \label{eq:StauRGE}
	16\pi^2 \, \mu\frac{\D}{\D\mu} m^2_{\tilde\tau_\mathrm{R}} =
	4 y_\tau^2 \, (m^2_{H_d} + m^2_{\tilde\tau_\mathrm{L}} +
	 m^2_{\tilde\tau_\mathrm{R}}) +
	4 a_\tau^2 -
	\frac{24}{5} g_1^2 M_1^2 \;,
\end{equation}
and of the other parameters of the theory.
In order to demonstrate the typical order of magnitude,
\Figref{fig:RatioMStauMBino} shows the ratio of the
$\tilde\tau_\mathrm{R}$ and the bino mass at the electroweak scale
calculated in a crude approximation: Mixing has been neglected, and only
the terms proportional to $M_1^2$ and
$m^2_{\tilde\tau_\mathrm{R}}$ in \Eqref{eq:StauRGE}
have been taken into account.  We have chosen
$\tan\beta=10$, $m_{1/2}(M_c)=400\GeV$ 
and $m^2_{H_u}(M_c)=m^2_{H_d}(M_c)=0$.
From the figure we see that in this case
the $\tilde\tau$ is typically lighter than the $\tilde B$, but not
by a large factor, so that it can easily be heavier than the gravitino.
A comparison with the full two-loop calculations 
\cite{Belanger:2005jk,*Allanach:2003jw} shows that the accuracy of the
approximation is reasonable for small $\tan\beta$ (and
$m^2_{H_u}=m^2_{H_d}=0$ at $M_c$) but quickly worsens for values larger
than about 20.  Then, the neglected effects are important and the actual
stau mass becomes significantly lighter than the estimate.  We have also
neglected the modifications to the running above the GUT scale for large
$M_c$, which tend to make the stau heavier
\cite{Schmaltz:2000gy,*Schmaltz:2000ei}.

\section{Conclusions}
We have studied constraints on gaugino and gravitino masses in models
with gaugino mediated SUSY breaking.  Based on naive dimensional
analysis, we have derived an upper bound on the coupling responsible for
gaugino masses.  This leads to a lower bound on the mass ratio
$m_{3/2}/m_{1/2}$, which allows for a gravitino LSP in a large domain of
parameter space.  Some regions in parameter space are now allowed
that were previously discarded in order to avoid a stau LSP.  In
particular, the compactification scale can coincide with the GUT scale
even in the minimal scenario with vanishing Higgs soft masses.  Fixing
the gaugino mass, one can translate the result for the mass ratio into a
lower bound on the gravitino mass.  For a gluino mass of $1\TeV$, we
find $m_{3/2} \gtrsim 10\GeV$.

A gravitino LSP can be naturally accompanied by a stau NLSP.  Long-lived
staus will then be observed at future colliders, and in their decays the
gravitino may be discovered.

\section*{Acknowledgments}
We would like to thank Adam Falkowski, J\"org J\"ackel, Hyun Min Lee,
Michael Ratz, Kai Schmidt-Hoberg, Michele Trapletti and Alexander
Westphal for helpful discussions.
This work has been supported by the ``Impuls- und Vernetzungsfonds'' of
the Helmholtz Association, contract number VH-NG-006.

\frenchspacing
\bibliography{GauginoMediation}
\bibliographystyle{ArXivmcite}

\end{document}